# Ultrafast synthesis of Pb-doped BiCuSeO oxyselenides by high-energy ball milling

Andrei Novitskii[1,*], Illia Serhiienko[1], Evgeny Kolesnikov[1], Elena Zakharova[1], Andrei Voronin[1], Vladimir Khovaylo[1,2]

[1]National University of Science and Technology MISIS, Moscow 119049, Russia

[2]National Research South Ural State University, Chelyabinsk 454080, Russia

**Abstract**

We report on the direct ultrafast synthesis of Pb-doped BiCuSeO oxyselenides by high-energy ball milling. We show that, contrary to the mechanical alloying sintering routes used in the previous reports which require 7 – 13 hours to fabricate the BiCuSeO phase in powder form, this synthesis technique enables us to obtain pure phase materials at room temperature under air with milling time less than 60 minutes.

**1. Introduction**

Thermoelectric materials, capable of directly converting heat energy to electrical energy based on the Seebeck effect or providing solid-state electronic cooling based on the Peltier effects. The energy conversion efficiency

$$\eta = \frac{T_h - T_c}{T_h} \frac{\sqrt{1+Z\overline{\overline{T}}} - 1}{\sqrt{1+Z\overline{\overline{T}}} + \frac{T_c}{T_h}} \qquad (1)$$

of thermoelectric devices, in general, is determined by the temperature difference between the hot side, $T_h$, and the cold side, $T_c$, of the thermoelectric legs, and by the constituting materials' dimensionless figure of merit $zT = S^2\sigma/\kappa$, where $T$ is the absolute temperature, $S$ is the Seebeck coefficient, $\sigma$ is the electrical conductivity, $\kappa$ is the total thermal conductivity and $Z\overline{T}$ is the average $zT$ value between $T_c$ and $T_h$ [1]. The low conversion efficiency, scarcity of constituents and high cost of state-of-art thermoelectric materials hinder the large-scale application of thermoelectric technology. Besides, vast majority of conventional thermoelectric materials may be oxidised, vaporised, decomposed or even melt when operating at high temperatures. Therefore, complex metal oxides composed of earth-abundant, light and cheap elements are expected to play a vital role in thermoelectric application due to their long-term thermal stability, nontoxicity and relatively high thermoelectric performance.

Since BiCuSeO oxyselenides were considered as potential thermoelectric materials in 2010, they have been paid considerable attention, and they have been extensively studied [2,3]. The highest figure of merit, $zT$, of 1.2 – 1.5 has been achieved in BiCuSeO by utilising the approaches such as carrier concentration optimisation [4–6], band structure engineering [7,8], modulation doping [9], texturation [10], etc. However, not only a large figure of merit is essential for industrial applications. Another critical issue is to develop an easily scalable, industrially friendly and straightforward synthesis technique, avoiding long thermal treatments at high temperatures under a controlled atmosphere over a long period of time. Ball milling technique is a simple, cost-effective and widely used approach for powder pulverisation that can also be used for mechanical alloying and materials synthesis [11–13]. Moreover, considering that during ball milling process temperature does not exceed 200 °C, it allows maintaining the stoichiometry close to nominal one even for compounds containing volatile elements whereas this is a challenge in the case of such conventional techniques as two-step solid-state reaction (SSR) [14].

---

[*] Corresponding author. Tel.: +7 916 504 2643
E-mail address: novitskiy@misis.ru (A. Novitskii)

In this work, we performed the high-energy ball milling (HBM) procedure for the pristine and Pb-doped $Bi_{1-x}Pb_xCuSeO$ ($x$ = 0, 0.02, 0.04, 0.06 and 0.08). We investigate the BiCuSeO phase formation mechanism during HBM and report on the direct ultrafast synthesis of these materials at low temperature under air atmosphere.

**2. Experimental section**

Stoichiometric mixture of high purity commercial powders Bi, $Bi_2O_3$, Se, Cu, and PbO as raw materials was subjected to HBM in air atmosphere. Milling was performed at 580 rpm for several milling times in a high-speed shimmy ball mill SFM-1 (MTI Corp., USA) with a powder-to-balls mass ratio of 1:40, alumina balls (⌀ = 10 mm) and vials (500 ml). The total masses of precursor powders used in the experiments were 5 grams. X-ray diffraction (XRD) analysis was performed by a DRON-3 diffractometer (IC Bourevestnik, USSR) using $CoK_α$ radiation (λ = 1.7903 Å). The Rietveld refinements were performed using the FullProf Suite [15]. The microstructure of the powders was observed by scanning electron microscopy (SEM, Vega 3 SB, Tescan, Czech Republic) and the chemical compositions were analyzed by energy-dispersive X-ray spectroscopy (EDS). Specific surface area, $A$, of the synthesised powders was determined from the low-temperature adsorption isotherms measurements performed by the Brunauere-Emmette-Teller (BET) method using a Nova 1200e analyser (Quantachrome Instruments, USA). The effective average particle size was calculated as $D = 6\rho^{-1}A^{-1}$ (where $\rho$ = 8.910 g·cm$^{-3}$ is the density of BiCuSeO) in the assumption of identical size and spherical shape for all particles.

**3. Results and discussion**

Considering the phase formation mechanism observed by XRD and EDS analyses (see Figs. 1 and 2), the BiCuSeO single-phase (PDF#48-0296) was produced after HBM for 60 minutes, which is five times faster compared to the previous reports [12,13,16]. The main peak of the 1111 phase, located at $2\theta ≈ 35°$, started to be clearly seen when the milling time reached 10 minutes (see Fig. 1a). With further increase of the milling time, intensity of the BiCuSeO phase Bragg peaks continuously grows, while the peaks corresponding to the starting phases fade and no other peaks than that of the main BiCuSeO phase were observed when a milling time of more than 30 minutes was used. Moreover, according to EDS data, the BiCuSeO phase already existed even after 5 minutes of HBM along with $Cu_{2-x}Se$ (where $x$ = 0.8 – 1) and $Bi_2O_3$ phases (see Fig. 1b). However, negligible contamination of the powders by the grinding media was also observed by EDS (see Figs. 2c, 2e). The Rietveld refinement was performed for samples with 30-, 60- and 90-minutes milling time. Detailed information on the refinement could be found in Electronic Supplementary Information (ESI). The corresponding lattice parameters are presented in Table 1. These values are in good agreement and slightly lower in comparison with those reported for samples synthesised using conventional solid-state route [5] and mechanical alloying (MA) [13]. Such difference in the lattice constants could be associated with the much smaller grain size of the samples obtained by HBM (see Table 1). Furthermore, the large strain induced by the high-energy non-equilibrium synthesis route may be a reason for the lattice parameters decrease.

**Table 1.** The experimental lattice constants and effective average particles size of BiCuSeO powders obtained with different milling time

| Milling time, min | Lattice constants | | Surface area, m$^2$·g$^{-1}$ | Effective average particles size, nm |
|---|---|---|---|---|
| | $a$, Å | $c$, Å | | |
| 30 | 3.924(1) | 8.963(3) | 1.213 | 550 ± 50 |
| 60 | 3.924(2) | 8.984(5) | 1.862 | 360 ± 40 |
| 90 | 3.920(1) | 8.996(5) | 2.724 | 240 ± 30 |

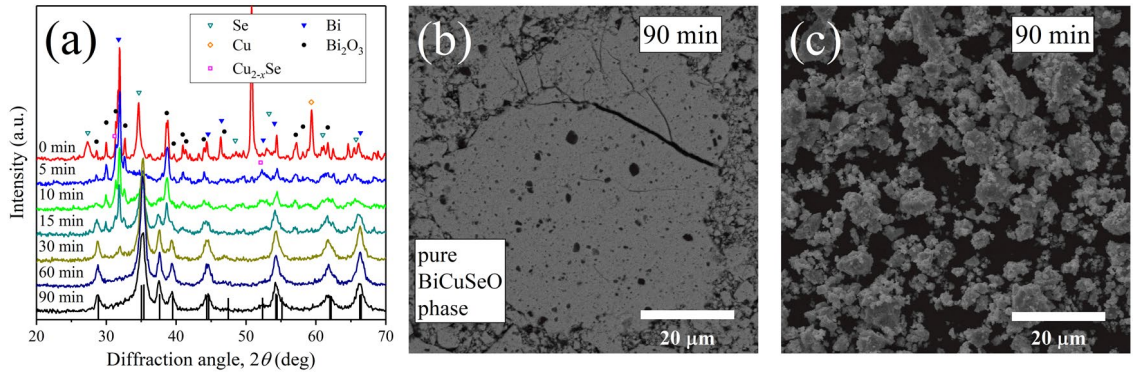

**Fig. 1.** (a) XRD patterns evolution for the starting mixture of the raw materials with nominal composition of BiCuSeO with increasing milling time, SEM images of (b) cold-pressed BiCuSeO powder and (c) powder particles after milling for 90 minutes

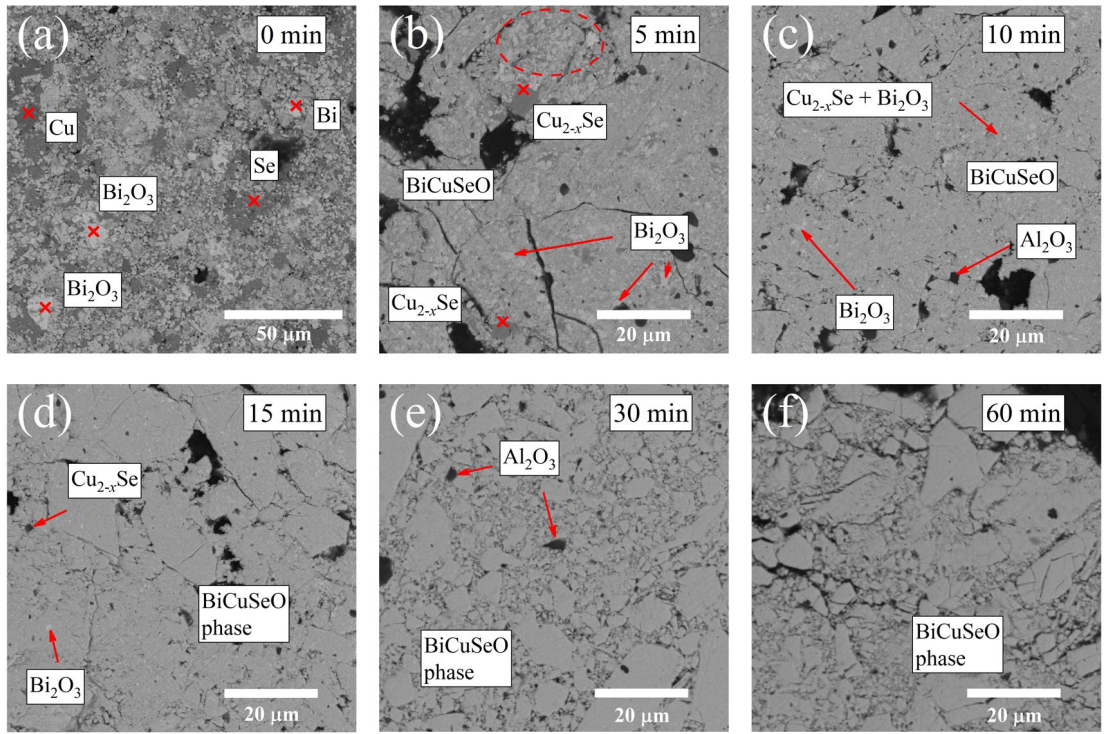

**Fig. 2.** SEM images of solid-state pressed powders after different milling time

According to the XRD and EDS data, the BiCuSeO phase formation mechanism during high-energy ball milling can be suggested as the following reactions:

$$(2 - x)\text{Cu} + \text{Se} = \text{Cu}_{2-x}\text{Se}, \quad (2)$$

while $Bi_2O_3$ doesn't react with Bi, Se or Cu. With further increase in the milling time, the main BiCuSeO phase is formed *via* reaction of bismuth, bismuth oxide and copper selenide,

$$3\text{Cu}_{2-x}\text{Se} + \text{Bi} + \text{Bi}_2\text{O}_3 = 3\text{BiCuSeO} \quad (3)$$

It seems that the energy applied to the material during HBM is so high that even at room temperatures (though, some heating is possible during the milling process) BiCuSeO phase formation occurred without intermediate $Bi_2SeO_2$ phase formation as is the case for BiCuSeO formation through the self-propagating high-temperature synthesis (SHS) or SSR [17].

It has been widely reported by many groups that one of the most effective way to enhance the BiCuSeO thermoelectric performance is carrier concentration optimisation through doping on Bi site. Therefore, in the context of this work it is necessary to show that doped samples can also be fabricated by HBM. Since Pb is one of the most effective dopants for the BiCuSeO system and Pb doping is beneficial for both carrier concentration increase and thermal conductivity reduction by point defect scattering [18], we synthesised $Bi_{1-x}Pb_xCuSeO$ ($x$ = 0, 0.02, 0.04, 0.06, 0.08) series using milling time of 90 minutes. No impurity phase was observed within XRD detection limit and all the peaks were attributed to the BiCuSeO phase (see Fig. 3). Moreover, a clear peak shift can be observed in the Pb-doped samples as compared to the pristine specimen, which corresponds to a linear increase of the lattice parameters as displayed in Fig. 3b for $Bi_{1-x}Pb_xCuSeO$ ($x$ = 0, 0.02, 0.04, 0.06 and 0.08) series and can be described by Vegard's law [19]. Thus, it was shown that single phase heavily doped BiCuSeO based ceramics can be synthesised by HBM.

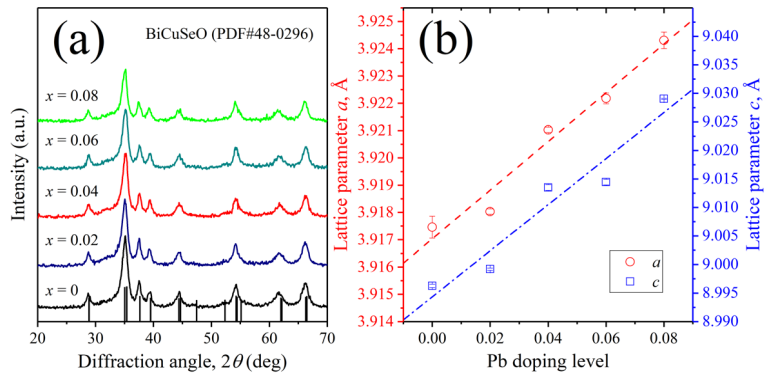

**Fig. 3.** (a) X-ray diffraction patterns and (b) lattice parameters of the $Bi_{1-x}Pb_xCuSeO$ ($x$ = 0, 0.02, 0.04, 0.06 and 0.08) series

### 4. Conclusions

It was shown, that BiCuSeO-based thermoelectric materials can be fabricated by high-energy ball milling in less than 90 min, resulting in single-phase powders with particle size ~240 nm. Moreover, this sintering technique allows maintaining the stoichiometry since all steps can be performed at room temperatures for a short time and can be easily scaled for commercial production. It is noteworthy that the contribution of the hierarchic structural features including point defects, nanograins and grain boundaries that may originate from the nonequilibrium HBM process could play a vital role for the electrical and the thermal transport behaviour of bulk BiCuSeO oxyselenides [2,20,21], which will certainly motivate future experimental studies.

### Acknowledgements

The study was carried out with financial support from the Russian Science Foundation (project No. 19-79-10282). VK acknowledges Act 211 Government of the Russian Federation, contract No. 02.A03.21.0011.

# Ultrafast synthesis of Pb-doped BiCuSeO oxyselenides by high-energy ball milling

Andrei Novitskii[1,*], Illia Serhiienko[1], Evgeny Kolesnikov[1], Elena Zakharova[1], Andrei Voronin[1], Vladimir Khovaylo[1,2]

[1]National University of Science and Technology MISIS, Moscow 119049, Russia

[2]National Research South Ural State University, Chelyabinsk 454080, Russia

[*]e-mail: novitskiy@misis.ru

**Rietveld refinement of the starting mixture with nominal composition of BiCuSeO with different milling time**

Figure S1 displays the Rietveld refinement of the BiCuSeO samples prepared with 30 min (Fig. S1a), 60 min (Fig. S1b) and 90 min milling time (Fig. S1c). The peak marked with green triangle corresponds to $Bi_2O_3$ phase. All the samples were milled at 580 rpm, with a powder-to-balls ratio of 1:40.

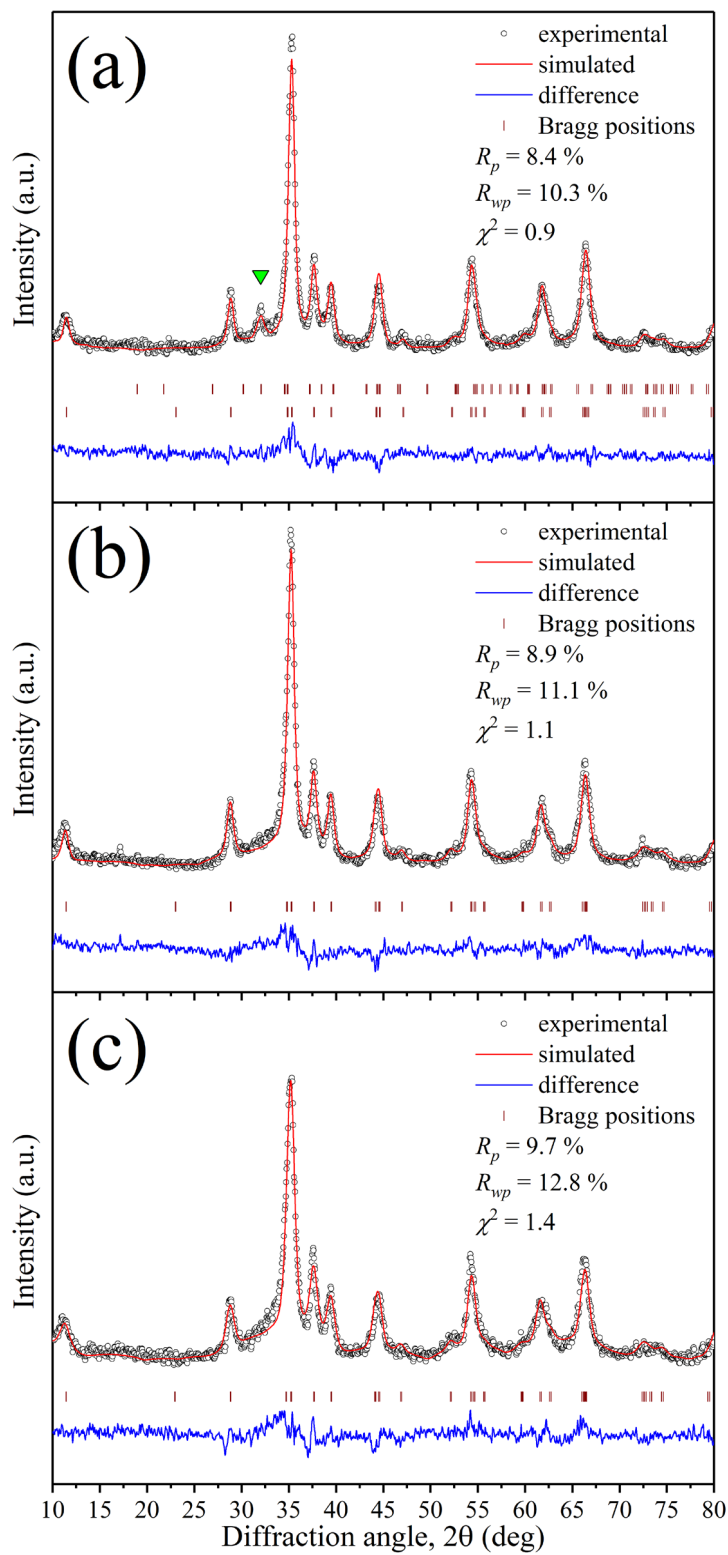

Figure S1 – Rietveld refinement for the starting mixture of the raw materials with nominal composition of BiCuSeO with milling time of (a) 30 min, (b) 60 min and (c) 90 min

**Rietveld refinement of the Pb-doped BiCuSeO samples (milling time = 90 min)**

Figure S2 shows the results of Rietveld fitting for the Pb-doped $Bi_{1-x}Pb_xCuSeO$ ($x$ = 0, 0.02, 0.04, 0.06 and 0.08) series. For all the patterns: $R_p \leq 8.1$ %, $R_{wp} \leq 10.9$ %, $\chi^2 \leq 1.4$.

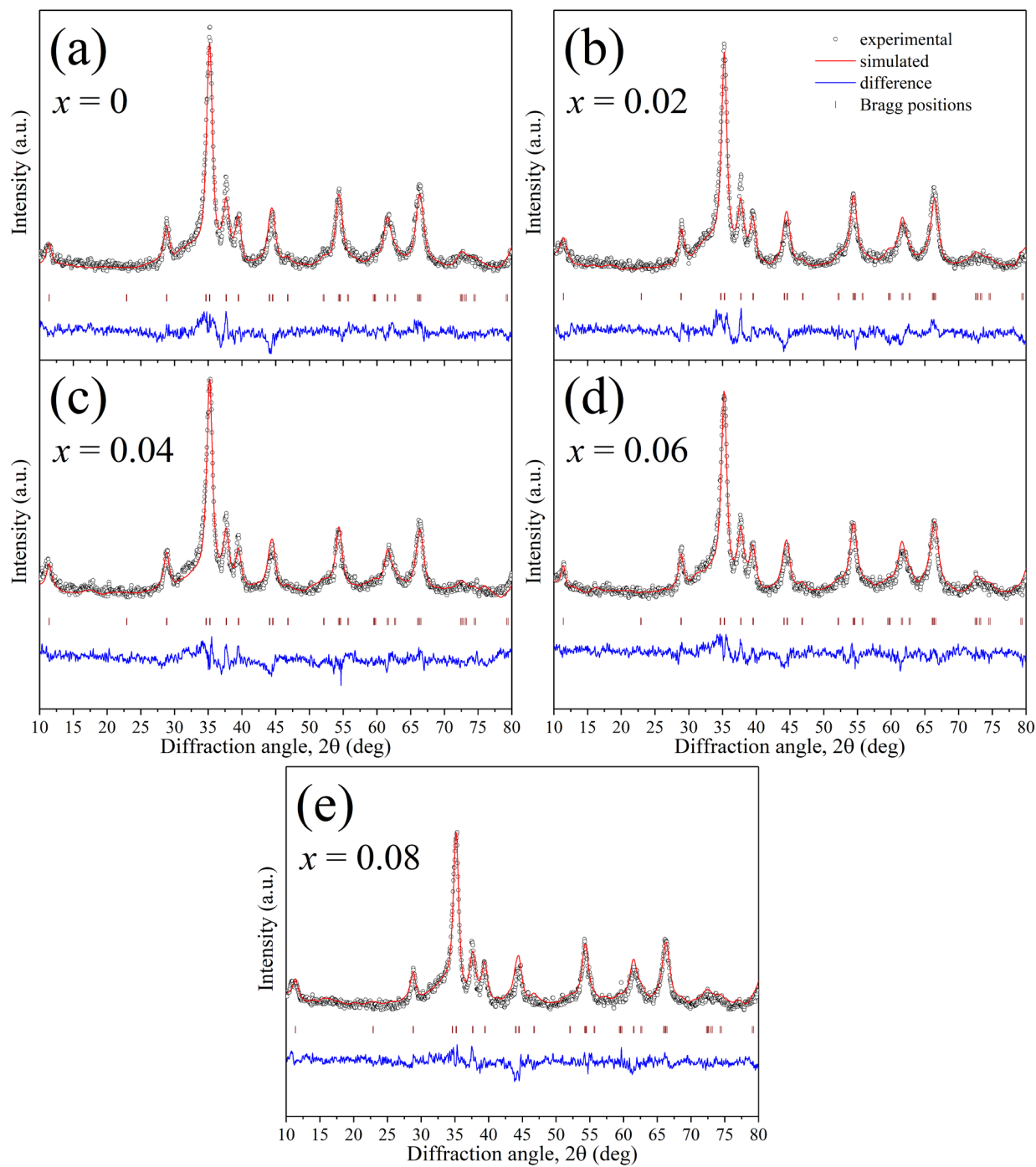

Figure S2 – Rietveld refinement of powder XRD pattern of $Bi_{1-x}Pb_xCuSeO$ ($x$ = 0, 0.02, 0.04, 0.06 and 0.08)